\documentstyle[preprint,aps]{revtex}
\begin{document}

\draft
\title{Inflation From Symmetry Breaking Below the Planck Scale}
\author{William H. Kinney\thanks{kinney@colorado.edu} and K.T.
Mahanthappa\thanks{ktm@verb.colorado.edu}}
\address{Dept of Physics, Campus Box 390}
\address{University of Colorado, Boulder CO 80309-0390}
\author{COLO-HEP-365, hep-ph/9511460}
\author{Submitted to Physical Review Letters}
\date{November 28, 1995}
\maketitle

\begin{abstract}
We investigate general scalar field potentials \hbox{$V\left(\phi\right)$}
for inflationary cosmology arising from spontaneous symmetry breaking. We
find that potentials which are dominated by terms of order $\phi^m$ with
\hbox{$m > 2$} can satisfy observational constraints at an arbitrary symmetry
breaking scale. Of particular interest, the spectral index of density
fluctuations is shown to be independent of the specific form of the
potential, depending only on the order $m$ of the lowest non-vanishing
derivative of $V(\phi)$ near its maximum. The results of a model with a
broken ${\rm SO(3)}$ symmetry illustrate these features.
\end{abstract}


Scalar field potentials arising from spontaneous symmetry breaking can in
general be characterized by the presence of a ``false'' vacuum, an unstable
or metastable equilibrium with nonzero vacuum energy density, and a physical
vacuum, for which the vacuum expectation value of the scalar field $\phi$ is
nonzero. At the physical vacuum, the potential has a stable minimum where the
vacuum energy density is defined to vanish. Take a potential
\hbox{$V\left(\phi\right)$} described by a symmetry breaking scale $v$ and a
vacuum energy density $\Lambda^4$:
\begin{equation}
V\left(\phi\right) = \Lambda^4 f\left({\phi \over
v}\right).\label{dimensionlessf}
\end{equation}
We restrict ourselves to potentials for ``new'' inflation, where the false
vacuum is an unstable equilibrium, and inflation takes place during a period
of {\it slow-roll}, in which the acceleration of the scalar field $\phi$ is
negligible, \hbox{$\ddot\phi \sim 0$}. We take the first derivative of the
potential to be zero at the origin and the minimum to be at \hbox{$\phi =
\phi_{min} \sim v$}, and the function $f$ satisfies \hbox{$f'\left(0\right) =
f'\left(\phi_{min}\right) = 0,\ f\left(0\right) = 1,\
f\left(\phi_{min}\right) = 0$}.

A scalar field initially at some value $\phi$ near the false vacuum evolves
to the minimum of the potential, where it oscillates and decays into other
particles ({\it reheating}). Inflation ends and reheating commences at a
field value $\phi_f$, where the first order parameter
\hbox{$\epsilon\left(\phi_f\right)$} is unity\cite{copeland93}:
\begin{eqnarray}
\epsilon\left(\phi_f\right) \equiv&& {m_{Pl}^2 \over 16 \pi}
\left({V'\left(\phi_f\right) \over V\left(\phi_f\right)}\right)^2 \equiv 1\cr
=&& {1 \over 16 \pi} \left({m_{Pl} \over v}\right)^2 \left({f'\left(\phi_f /
v\right) \over f\left(\phi_f / v\right)}\right)^2,\label{genericepsilon}
\end{eqnarray}
where \hbox{$m_{Pl} \sim 10^{19}\,{\rm GeV}$} is the Planck scale. An
inflationary phase is characterized by \hbox{$\epsilon < 1$}: here
\hbox{$\epsilon\left(\phi = 0\right) = 0$} by construction. If
\hbox{$\epsilon\left(\phi\right)$} is everywhere increasing on the range
\hbox{$0 \leq \phi < \phi_{min}$}, there is a unique field value $\phi_f$ at
which inflation ends, where \hbox{$\epsilon\left(\phi_f\right) \equiv 1$} and
\hbox{$\epsilon\left(\phi < \phi_f\right) < 1$}.
We are particularly interested in cases where the symmetry breaking takes
place well below the Planck scale, \hbox{$v \ll m_{Pl}$}. Noting from
(\ref{genericepsilon}) that \hbox{$\epsilon \propto \left(m_{Pl} /
v\right)^2$}, the field value $\phi_f$ at which inflation ends is small for
\hbox{$v \ll m_{Pl}$}, and we need only consider the behavior of the
potential near the origin. We can Taylor expand $V\left(\phi\right)$ about
the origin:
\begin{equation}
V\left(\phi\right) = V\left(0\right) + {1 \over m!} {d^m V \over d \phi^m}
\biggr|_{\phi =  0} \phi^m + \cdots,\label{generalTaylorexpansion}
\end{equation}
where \hbox{$V'\left(0\right) \equiv 0$}, and $m$ is the order of the lowest
non-vanishing derivative at the origin. For cases in which the origin is a
maximum of the potential, $m$ must be even, and \hbox{$d^m V / d \phi^m <
0$}. For $m$ odd, the origin is at a saddle point, and we can define the
positive $\phi$ direction to be such that \hbox{$d^m V / d \phi^m < 0$}.
Models dominated by terms of order $m = 2$ are frequently discussed in the
literature. Here we consider potentials for which the second derivative
vanishes at the origin, \hbox{$m > 2$}. It is to be expected that for most
potentials arising from spontaneous symmetry breaking, inflation will take
place near an unstable maximum and $m$ will be even, but this is an
unnecessarily strict condition for the purpose of a general analysis. The
potential can be written in the form
\begin{equation}
V\left(\phi\right) = \Lambda^4 \left[1 - {1 \over m} \left({\phi \over
\mu}\right)^m + \cdots\right],\label{generalV}
\end{equation}
so that for \hbox{$\left(\phi / \mu\right) \ll 1$}, the potential is
dominated by terms of order \hbox{$\left(\phi / \mu\right)^m$}. The vacuum
energy density is \hbox{$\Lambda^4 \equiv V\left(0\right)$}, and \hbox{$\mu
\propto v$} is an effective symmetry breaking scale defined by
\begin{equation}
\mu \equiv \left({(m - 1)! V\left(\phi\right) \over \left|d^m V / d
\phi^m\right|}\right)^{1 / m}\Biggr|_{\phi = 0} = v \left({(m - 1)!
f\left(x\right) \over \left|d^m f / d x^m\right|}\right)^{1 / m}\Biggr|_{x =
0}.\label{effectivemassscale}
\end{equation}
The first order inflationary parameter $\epsilon$ is given by
\begin{eqnarray}
\epsilon\left(\phi\right) =&& {1 \over 16 \pi} \left(m_{Pl} \over
\mu\right)^2 \left({\left(\phi / \mu\right)^{m - 1} \over 1 - \left(1 /
m\right) \left(\phi / \mu\right)^m}\right)^2\cr
\simeq&& {1 \over 16 \pi} \left(m_{Pl} \over \mu\right)^2 \left(\phi \over
\mu\right)^{2 (m - 1)}.
\end{eqnarray}
Taking \hbox{$\epsilon\left(\phi_f\right) \equiv 1$}, we have for $\phi_f$
\begin{equation}
\left(\phi_f \over \mu\right) = \left[\sqrt{16 \pi} \left(\mu \over
m_{Pl}\right)\right]^{1 / \left(m - 1\right)}.\label{phifho}
\end{equation}
The number of e-folds of inflation which occur when the field evolves from
$\phi$ to $\phi_f$ is\cite{parsons95}
\begin{eqnarray}
N\left(\phi\right) =&& {8 \pi \over m_{Pl}^2}
\int_{\phi_f}^{\phi}{{V\left(\phi'\right) \over
V'\left(\phi'\right)}\,d\phi'}\cr
=&& - 8 \pi \left(\mu \over m_{Pl}\right)^2 \int_{\phi_f / \mu}^{\phi / \mu}
{{1 - x^m /m \over x^{m - 1}}dx}\cr
\simeq&& 8 \pi \left(\mu \over m_{Pl}\right)^2 \left(1 \over m - 2\right)
\left[\left(\mu \over \phi\right)^{m - 2} - \left(\mu \over \phi_f\right)^{m
- 2}\right].\label{Nwithfinal}
\end{eqnarray}
Smoothness on scales comparable to the current horizon size requires \hbox{$N
\geq 60$}, which places an upper limit on the initial field value \hbox{$\phi
\leq \phi_{60}$}, where \hbox{$N\left(\phi_{60}\right) \equiv 60$}.
Substituting (\ref{phifho}) for $\phi_f$, we then have for $\phi_{60}$,
\begin{equation}
\left(\phi_{60} \over \mu\right) = \left\lbrace{15 (m - 2) \over 2 \pi}
\left(m_{Pl} \over \mu\right)^2 + \left[{1 \over \sqrt{16 \pi}} \left(m_{Pl}
\over \mu\right)\right]^{(m - 2) / (m - 1)}\right\rbrace^{-1 / (m - 2)}.
\end{equation}
Since \hbox{$\left(m - 2\right) / \left(m - 1\right) < 1$}, the
\hbox{$\left(m_{Pl} / \mu\right)^2$} term dominates, and we have the result
that $\phi_{60}$ is to a good approximation {\it independent} of $\phi_f$ for
\hbox{$\mu \ll m_{Pl}$}:
\begin{equation}
\left(\phi_{60} \over \mu\right) \simeq \left[{2 \pi \over 15 (m - 2)}
\left(\mu \over m_{Pl}\right)^2\right]^{1 / (m - 2)}.\label{phiiho}
\end{equation}
This independence will be of importance when we consider the consistency of
the slow-roll approximation. Quantum fluctuations in the inflaton field
produce density fluctuations on scales of current astrophysical interest when
\hbox{$\phi \sim \phi_{60}$}. The scalar density fluctuation amplitude
produced during inflation is given by:
\begin{eqnarray}
\delta =&& \sqrt{2 \over \pi} {\left[V\left(\phi_{60}\right)\right]^{3/2}
\over m_{Pl}^3 V'\left(\phi_{60}\right)}\cr
=&& \sqrt{2 \over \pi} \left({\Lambda^2 \mu \over m_{Pl}^3}\right) {\left[1 -
\left(1 / m\right) \left(\phi_{60} / \mu\right)^m\right]^{(3/2)} \over
\left(\phi_{60} / \mu\right)^{m - 1}}\cr
\simeq&& \sqrt{2 \over \pi}\left({\mu \over m_{Pl}}\right)^3 \left({\Lambda
\over \mu}\right)^2 \left({\mu \over \phi_{60}}\right)^{m - 1}.
\end{eqnarray}
Substituting $\phi_{60}$ from (\ref{phiiho}), we have
\begin{equation}
\delta = \sqrt{2 \over \pi} \left(15 (m - 2) \over 2 \pi\right)^{(m - 1) / (m
- 2)} \left(\Lambda \over \mu\right)^2 \left(\mu \over m_{Pl}\right)^{(m - 4)
/ (m - 2)}.\label{deltaho}
\end{equation}
For the case $m = 4$, the density fluctuation amplitude is independent of
\hbox{$\left(\mu / m_{Pl}\right)$}. For \hbox{$m > 4$}, $\delta$ decreases
with decreasing \hbox{$\left(\mu / m_{Pl}\right)$} -- production of density
fluctuations is suppressed at low scale. We can constrain the scale $\Lambda$
by using the Cosmic Background Explorer (COBE) measurement, \hbox{$\delta
\simeq 10^{-5}$}\cite{smoot92,wright92}:
\begin{equation}
\left({\Lambda \over \mu}\right)^2 = \delta \sqrt{\pi \over 2} \left(2 \pi
\over 15 (m - 2)\right)^{(m - 1) / (m - 2)} \left(m_{Pl} \over \mu
\right)^{(m - 4) / (m - 2)},\label{Lambdaho}
\end{equation}
and the constraint requires no fine-tuning of constants. Inflation is
consistent only if $\phi_{60}$ is greater than the magnitude of quantum
fluctuations on the scale of the horizon size $\phi_q$, where
\begin{eqnarray}
\left(\phi_q \over \mu\right) =&& {1 \over 2 \pi \mu} \sqrt{{8 \pi \over 3
m_{Pl}^2} V\left(0\right)}\cr
=&& \sqrt{2 \over 3 \pi} \left(\mu \over m_{Pl}\right) \left(\Lambda \over
\mu\right)^2\cr
=&& {\delta \over \sqrt{3}} \left(2 \pi \over 15 (m - 2)\right)^{(m - 1) / (m
- 2)} \left(\mu \over m_{Pl}\right)^{2 / (m - 2)},
\end{eqnarray}
and the consistency condition $\left(\phi_q / \phi_{60}\right) < 1$ is
satisfied independent of $\left(\mu / m_{Pl}\right)$:
\begin{equation}
\left(\phi_q \over \phi_{60}\right) = {2 \pi \delta \over 15 \sqrt{3} (m -
2)}.
\end{equation}
The spectral index of density fluctuations, $n_s$, is given in terms of the
slow-roll parameters $\epsilon$ and $\eta$\cite{stewart93}:
\begin{equation}
n_s \simeq 1 - 4 \epsilon\left(\phi_{60}\right) + 2
\eta\left(\phi_{60}\right)\label{scalarindex},
\end{equation}
where the second order slow-roll parameter $\eta$ is defined to be:
\begin{equation}\eta\left(\phi\right) \equiv {m_{Pl}^2 \over 8 \pi}
\left[{V''\left(\phi\right) \over V\left(\phi\right)} - {1 \over 2}
\left({V'\left(\phi\right) \over V\left(\phi\right)}\right)^2\right].
\end{equation}
For the potential (\ref{generalV}), \hbox{$V'\left(\phi_{60}\right) \sim 0$},
and
\begin{eqnarray}
n_s \simeq&& 1 + {m_{Pl}^2 \over 4 \pi} {V''\left(\phi_{60}\right) \over
V\left(\phi_{60}\right)}\cr
\simeq&& 1 - {m - 1 \over 4 \pi} \left(m_{Pl} \over \mu\right)^2
\left(\phi_{60} \over \mu\right)^{m - 2}\cr
=&& 1 - \left(1 \over 30\right) {m - 1 \over m - 2},\label{spectralindex}
\end{eqnarray}
and we have the rather surprising result that for any \hbox{$m > 2$}, the
scalar spectral index is independent of any characteristic of the potential
except the order of the lowest non-vanishing derivative at the origin. The
scalar spectral index is nearly scale invariant, with \hbox{$0.93 < n_s <
0.97$} for all values of $m$. Thus we have the result that potentials
characterized by \hbox{$m > 2$} can naturally satisfy observational
constraints for {\it any} effective symmetry breaking scale $\mu$, where
$\mu$ is proportional to the vacuum expectation value of the scalar field
$\phi$. The case \hbox{$m = 4$} is particularly well-behaved, with both
$\delta$ and $n_s$ independent of \hbox{$\left(\mu / m_{Pl}\right)$} -- the
Planck scale drops out of the constraints altogether.

One apparent difficulty with this class of potentials, however, is that the
second order slow-roll parameter $\left|\eta\right|$ becomes large for
\hbox{$\phi \ll \phi_f$}, so that the slow-roll approximation is invalid over
much of the range at which inflation is taking place. Inflation ends at
$\phi_f$ given by (\ref{phifho}), but {\it slow-roll} ends at
\begin{eqnarray}
&&\left|\eta\left(\phi\right)\right| \simeq {m_{Pl}^2 \over 8 \pi}
\left|{V''\left(\phi\right) \over V\left(\phi\right)}\right| = 1,\cr
&&\left(\phi \over \mu\right) = \left[{8 \pi \over m - 1} \left(\mu \over
m_{Pl}\right)^2\right]^{1 / (m - 2)} \ll \left(\phi_f \over
\mu\right).\label{endslowroll}
\end{eqnarray}
However, from in equation (\ref{phiiho}),  $\phi_{60}$ is {\it independent}
of $\phi_f$, so that the breakdown of slow-roll has no effect, as long as
slow-roll is valid at the initial field value,
\hbox{$\left|\eta\left(\phi_{60}\right)\right| < 1$}. If we define $\phi_{60}$
to be 60 e-folds before the end of slow-roll as defined in
(\ref{endslowroll}), instead of the end of inflation proper, we have, using
(\ref{Nwithfinal}) for $N\left(\phi\right)$,
\begin{equation}
\left(\phi_{60} \over \mu\right) \simeq \left[{2 \pi \over 15 (m - 2)}
\left(\mu \over m_{Pl}\right)^2\right]^{1 / (m - 2)} \left(1 + {m - 1 \over
60 (m - 2)}\right)^{-{1 / (m - 2)}},
\end{equation}
which is a small correction to equation (\ref{phiiho}).

Inflationary potentials characterized by \hbox{$m > 2$} can arise in
physically well motivated contexts. The ``natural inflation''
scenario\cite{freese90}, in which inflation is driven by a pseudo
Nambu-Goldstone boson, is one such case. It can be shown that in a Lagrangian
with a broken ${\rm SO(3)}$ symmetry, gauge loop effects can generate an
effective potential for a Nambu-Goldstone mode $\theta$ dominated by terms of
order $\theta^4$ for small $\theta$\cite{kinney95}:
\begin{eqnarray}
V\left(\theta\right)= {3 v^4 \over 64 \pi^2} g^4 \left\lbrace
\sin^4\left(\theta \over v\right) \ln\left[g^2 \sin^2\left(\theta \over
v\right)\right] - \ln\left(g^2\right)\right\rbrace,\label{so3pot}
\end{eqnarray}
where $g < 1$ is a gauge coupling constant. Note that this potential does not
have a well-defined Taylor expansion at the origin, since the fourth
derivative has a logarithmic (infared) divergence as \hbox{$\theta
\rightarrow 0$}. However, the results derived here for the general potential
are valid to corrections of order \hbox{$\ln\left(v / m_{Pl}\right)$}, and
form a lowest order approximation to the exact results. In particular, the
expression (\ref{spectralindex}) for the spectral index of scalar density
fluctuations is exact, with \hbox{$n_s = 0.95$}. A lower bound on the
symmetry breaking scale $v$ can be obtained with the inclusion of fermions. A
detailed analysis will be presented in a forthcoming paper\cite{kinney95}.

\end{document}